\documentclass[aps,pre,reprint,amsmath,amssymb,superscriptaddress]{revtex4-2}

\usepackage{natbib}
\usepackage{graphicx}

\usepackage[hidelinks]{hyperref}

\newcommand{\enquote}[1]{``#1''}

\newcommand{\dotsep}{$\cdot$}

\newcommand{\X}{\ensuremath{\mathrm{X}}}
\newcommand{\A}{\ensuremath{\mathrm{A}}}
\newcommand{\B}{\ensuremath{\mathrm{B}}}
\renewcommand{\S}{\ensuremath{\mathrm{S}}}
\newcommand{\I}{\ensuremath{\mathrm{I}}}

\newcommand{\stack}[2]{\ensuremath{\genfrac{}{}{0pt}{}{#1}{#2}}}

\usepackage{enumitem}

\begin{document}


\title{Phase transitions for polyadic epidemic and voter models with multiscale groups}

\author{Pia Steinmeyer}
\thanks{These authors contributed equally to this work.}
\affiliation{Department of Mathematics, School of Computation, Information and Technology, Technical University of Munich, Boltzmannstraße 3, 85748 Garching bei München, Germany}
\author{Jan Mölter}
\thanks{These authors contributed equally to this work.}
\affiliation{Department of Mathematics, School of Computation, Information and Technology, Technical University of Munich, Boltzmannstraße 3, 85748 Garching bei München, Germany}
\author{Christian Kuehn}
\affiliation{Department of Mathematics, School of Computation, Information and Technology, Technical University of Munich, Boltzmannstraße 3, 85748 Garching bei München, Germany}
\affiliation{Munich Data Science Institute, Technical University of Munich, Walther-von-Dyck-Straße 10, 85748 Garching bei München, Germany}
\affiliation{Complexity Science Hub Vienna, Josefstädter Straße 39, 1080 Vienna, Austria}

\begin{abstract}
    Polyadic (or higher-order) interactions can significantly impact the dynamics of interacting particle systems. However, previous studies have often assumed group sizes to be relatively small. In this work, we examine the influence of multiscale polyadic group interactions, where some groups are small and others are very large. We consider two paradigmatic examples, an SIS-epidemic and the adaptive voter model. On the level of the mean field, we specifically discuss the impact of the multiscale polyadic interactions on equilibrium dynamics and phase transitions. For the SIS-epidemic model, we find a region of bistability that protects the disease-free state over a wide range beyond the classical epidemic threshold from a significant outbreak. For the adaptive voter model, we show that multiscale polyadic interactions can stabilize the network or increase the convergence rate to an unbiased equilibrium.

    \bigskip
    
\end{abstract}

\maketitle


\section{Introduction}

In many systems of populations of interacting individuals, groups, and interactions within these groups may play an integral role in collective dynamics. However, for a long time, these polyadic (or higher-order) interactions have been largely neglected in favor of only dyadic interactions, such as in the classical models of the spreading of a disease~\cite{keeling2005networks,pastorsatorras2015epidemic,serrano2006percolation,pastorsatorras2001epidemic,boguna2013nature,gross2006epidemic}, opinion and consensus formation~\cite{clifford1973model,holley1975ergodic,sood2005voter,holme2006nonequilibrium,kimura2008coevolutionary,zschaler2012adaptive,fernandezgracia2014voter, nardini2008whos}, or swarming~\cite{huepe2011adaptive,chen2016adaptive}.
Only recently have polyadic interactions been taken explicitly into account. From a modeling perspective, instead of classical networks or graphs, this entails considering the dynamics on higher-order structures such as hypergraphs or simplicial complexes, where the latter is a special case of the former~\cite{bick2023what,battiston2020networks,battiston2021physics,zhang2023higherorder,bianconi2021higherorder,petri2018simplicial,courtney2016generalized}. In particular, it has been shown that this can considerably change the overall dynamics~\cite{majhi2022dynamics}. For instance, the emergence of a first-order phase transition as opposed to a second-order on, and multistability can be caused by polyadic interactions~\cite{iacopini2019simplicial,skardal2019abrupt,kuehn2021universal,ferrazdearruda2023multistability,burgio2024triadic}. Similarly, in the adaptive voter model, polyadic interactions can accelerate consensus formation~\cite{horstmeyer2020adaptive,papanikolaou2022consensus} and might affect the type of phase transition~\cite{golovin2024polyadic}. In all these works, the main focus was on extending dyadic interactions to slightly polyadic cases such as triadic or tetradic interactions.  

In this work, we want to consider multiscale polyadic interactions, i.e., interactions through both small and comparatively large groups. The idea is that there is a significant gap between the relevant group sizes, leading to a multiscale problem. This is directly motivated by applications, e.g., in epidemic spreading or voter/opinion dynamics. Beyond interactions in small groups, there are crucial ones that occur on the large-scale or societal level, e.g., running a health-care system and hospitals in particular, conducting (medical) research, participation in social groups such as clubs, or influence by mass media. To this end, we focus on an epidemic and a voter model on a simplicial complex and hypergraph, respectively. In their classical form, these models (an epidemic model being a special case of a contact process) are paradigmatic and the main and simplest examples of interacting particle systems that have been extensively studied for several decades now~\cite{liggett1999stochastic,liggett2005interacting}. In fact, both are important models of social dynamics, accounting for invasion/spreading or competition respectively~\cite{lanchier2024stochastic,castellano2009statistical,durrett2010some}. In particular, rather than considering the stochastic models, we will focus on corresponding mean-field models and study how the presence of large groups affects their equilibrium dynamics and, specifically, phase transitions. We are going to show that major changes in terms of stability and convergence to equilibrium can be induced by multiscale polyadic interactions. In particular, this demonstrates the potential impact of such interactions in not only one but two important and widely studied models of social dynamics.

\section{Results}

Our main tool for studying interacting particle systems is a mean-field description in terms of network moments. These correspond to the expected number of certain network motifs~\cite{keeling1999effects,house2009motif,porter2016dynamical,kiss2015generalization}. With increasing order, examples of these moments are the expected number of nodes in state $\X$, the expected number of connected nodes in states $\X$ and $\X'$, or the expected number of triples in state $\X$, $\X'$, and $\X''$ denoted as $[\X]$, $[\X\X']$, and $[\X\X'\X'']$, respectively.
The evolution equations for any finite set of moments are generically not closed as they form an infinite hierarchy of evolution equations~\cite{house2009motif,sharkey2008deterministic,sharkey2011deterministic,taylor2012markovian,kuehn2016moment}. To break this hierarchy and reduce the number of evolution equations to an amenable size, the prevailing strategy is to apply moment closures that approximate a given moment of a certain order in terms of lower-order ones~\cite{kuehn2016moment,sharkey2008deterministic,sharkey2011deterministic,taylor2012markovian,pellis2015exact,wuyts2022meanfield}. Using standard moment closures, in the following, we will consider mean-field evolution equations that have been closed at the lowest nontrivial order.

In the models we consider in the following, we introduce polyadic interactions in two different ways. First, in the epidemic model, we define the dynamics on a higher-order structure from the start. In this case, this structure is a special kind of simplicial complex, as they have been utilized before the context of epidemic dynamics~\cite{iacopini2019simplicial}. Second, in the adaptive voter model, we start with dyadic dynamics and couple them to a higher-order structure in the background.

\subsection{Model I: Multiscale polyadic SIS-epidemics}
Social contagion models have been successfully used to describe the spread of behaviors~\cite{cao2017method,xiao2021evacuation,hill2010emotions,cohen2003efficient,chen2008finding,granell2013dynamical,liu2014controlling}, goods~\cite{illner2016sistype}, and (mis)information flows~\cite{maleki2021using,granell2013dynamical} or, perhaps most commonly studied, diseases within a population~\cite{keeling2005networks,barrat2008dynamical,pastorsatorras2015epidemic,kuperman2001small,pastorsatorras2001epidemic,gleeson2011highaccuracy,eguiluz2002epidemic,newman2005threshold}.
In the context of an epidemic, each individual of a population of size $N$ is assumed to be either susceptible ($\S$) or infected ($\I$). The disease spreads along links of the underlying social network that connects the population, from infected to susceptible individuals, while infected individuals recover independently, both at times given by a Poisson process. Depending on the ratio between infection and recovery rate, the number of infected individuals either vanishes or not, in which case the epidemic can sustain itself.

To account for polyadic interactions, we consider an SIS-epidemic model on a simplicial $(N-1)$-network, i.e., the simplicial complex generated by an (abstract) $(N-1)$-simplex or, equivalently, the hypergraph generated by the power-set of a vertex set with $N$ elements. For this, the mean-field evolution equation in lowest order reads
\begin{equation*}
    \dot{[\I]} = - r_{0} [\I] + \sum_{g \geq 2}^{N} \sum_{\stack{\alpha = (\alpha_{\S},\alpha_{\I}) \in \mathbb{N}^{2}}{\alpha_{\S} +\alpha_{\I} = g}  }  (\alpha_{\S}  \lambda_{\alpha} - \alpha_{\I} r_{\alpha}) [\triangle \{  \S^{\alpha_{\S}} \I^{\alpha_{\I}} \} ].
\end{equation*}
where $[\triangle \{  \S^{\alpha_{\S}} \I^{\alpha_{\I}} \}]$ denotes the expected number of simplices with $\alpha_{\S}$ susceptible and $\alpha_{\I}$ infected individuals, extending the notation that we introduced above for network moments. With $\alpha = (\alpha_{\S}, \alpha_{\I})$, the parameters $r_{\alpha} \geq 0$ and $\lambda_{\alpha} \geq 0$ are recovery and infection rates, respectively, along faces with $\alpha_{\S}$ susceptible and $\alpha_{\I}$ infected individuals.
In particular, note that in the case where all $\lambda_{\alpha}$ and $r_{\alpha}$ vanish except $\lambda_{(1,1)}$ and $r_{0}$, this reduces to the well-known mean-field equation $\dot{[\I]} = - r_{0} [\I] + \lambda_{(1,1)} [ \triangle \{\S\I\}]$ with $[ \triangle \{\S\I\}] \equiv [\S\I]$~\cite{simon2011exact}.

In the following, we imagine a society where, in addition to pairwise connections between individuals that facilitate the spreading of an epidemic, some societal features or effects promote recovery beyond the independent, purely physiological recovery. In particular, we think of these effects as emanating from a rather large group of healthy (susceptible) individuals and received by a comparatively small number of infected individuals, such as when running hospitals, developing medicine, organizing care, etc.

For simplicity and the purpose of studying the interplay between infections through pairwise interactions and the presence of such large social components of some uniform size that promote recovery, we fix a group size $1 \ll g \ll N$ as well as a partition $g = g_{\S} + g_{\I}$ with $1 \leq g_{\S}, g_{\I} \leq g$ and suppose that all $\lambda_{\alpha}$ and $r_{\alpha}$ vanish except $\lambda_{(1,1)}$ and $r_{0}$ and $r_{(g_{\S}, g_{\I})}$. Consequently, the processes we consider are in addition to infections from an infected to a susceptible individual with rate $\lambda_{(1,1)}$ and independent recoveries of individuals with rate $r_{0}$ as in the standard models, also (simultaneous) recoveries of $g_{\I}$ infected individuals when connected in a group, i.e., in this case the face of a simplex, with $g_{\S}$ susceptible individuals with a rate $r_{(g_{\S}, g_{\I})}$. In particular, since we are interested in the case where the group is composed predominantly of susceptible and only a few infected, we further assume that $g_{\S} \gg g_{\I} \geq 1$.

With $\lambda := \lambda_{(1,1)}$ and $r := r_{(g_{\S}, g_{\I})}$ and measuring time in units such that $r_{0} = 1$, the above evolution equation reduces to
\begin{equation*}
    \dot{[\I]} = -[\I] + \lambda [\triangle\{\S\I\}] - r g_{\I} [ \triangle \{\S^{g_{\S}} \I^{g_{\I}} \} ].
    \label{eq:simplicial-multiscale-sis-epidemic}
\end{equation*}

In a simplicial $(N-1)$-network, one has the moment closure
$[\triangle\lbrace \S^{g_{\S}} \I^{g_{\I}} \rbrace] \approx [\S]^{g_{\S}} [\I]^{g_{\I}}$ (see Appendix~\ref{sec:appendix-simplicial-closure}) so that applying this and using the conservation relation $[\S] = N - [\I]$, the closed mean-field evolution equation is given as
\begin{equation*}
    \dot{[\I]} = -[\I] + \lambda  (N - [\I]) [\I] - r g_{\I} (N - [\I])^{g_{\S}} [\I]^{g_{\I}}.
    \label{eq:simplicial-multiscale-sis-epidemic_closed}
\end{equation*}

For $r = 0$, the above equation corresponds to the standard mean-field equation for the SIS-epidemic model with a (supercritical) transcritical bifurcation of the trivial, disease-free equilibrium $[\I] = 0$ at $\lambda N = 1$. Below this bifurcation point, this equilibrium is stable and it is the only equilibrium within the physical range, $0 \leq [\I] \leq N$. Above the bifurcation point, the disease-free equilibrium becomes unstable, whilst a nontrivial, endemic state arises as an equilibrium.

\begin{figure}[tbhp]
    \centering
    
    \includegraphics[page=1]{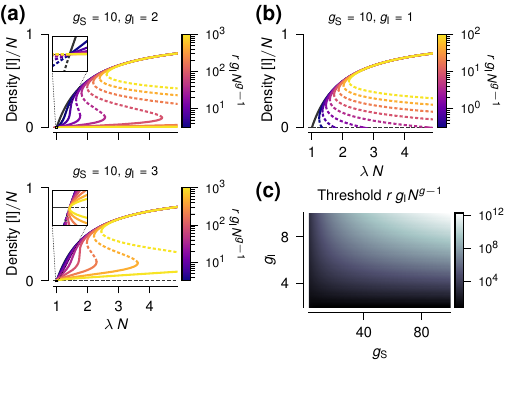}
    
    \vspace{-0.5cm}
    
    \caption{\textbf{The SIS-epidemic model with large recovery groups.} \textbf{(a)}: Phase diagrams for $g_{\I} = 2$ and $g_{\I} = 3$ ($g_{\I} > 3$ are qualitatively similar to $g_{\I} = 3$). The transcritical bifurcation occurs at $\lambda N = 1$. As $r g_{\I} N^{g-1}$ is increased, its slope becomes shallower in the case of $g_{\I} = 2$ while remaining constant in the case of $g_{\I} = 3$ (inset, showing a neighborhood of the bifurcation point within a radius of $3 \cdot 10^{-2}$), but remains supercritical. Globally, a double saddle-node bifurcation forms. \textbf{(b)}: Bifurcation diagrams for $g_{\I} = 1$. As $r g_{\I} N^{g-1}$ is increased, the transcritical bifurcation point shifts to the right, becomes shallower, and eventually turns subcritical. \textbf{(c)}: Threshold value of $r g_{\I} N^{g-1}$ as a function of the group composition above which the double saddle-node bifurcation forms. This value increases rapidly as $g_{\S}$ or $g_{\I}$ are increased (see Appendix~\ref{sec:appendix-SIS-double-saddle-node}).}
    \label{fig:sis-epidemic-model}
\end{figure}

For $r >0$, we observe that the system exhibits a transcritical bifurcation of the disease-free equilibrium $[\I] = 0$ 
at 
\begin{itemize}
    \item $\lambda N = 1$, for $g_{\I} \geq 2$, and
    \item $\lambda N = 1 + r N^{g_{\S}} = 1 + r g_{\I} N^{g - 1}$ for $g_{\I} = 1$.
\end{itemize}
More precisely, in a sufficiently small neighborhood of the bifurcation point, the system has the following normal forms
\begin{itemize}
    \item $\dot{[\I]} = (\lambda N - 1) [\I] - \frac{1}{N} [\I]^{2} + \mathcal{O}([\I]^{3})$, for $g_{\I}  \geq 3$,
    \item $\dot{[\I]} = (\lambda N - 1)[\I] - \frac{1}{N}(1 + 2 r N^{g_{\S}+1}) [\I]^{2} + \mathcal{O}([\I]^{3})$, for $g_{\I} = 2,$ and
    \item $\dot{[\I]} = (\lambda N - (1 + r N^{g_{\S}})) [\I] - \frac{1}{N}(1 - r (g_{\S}-1) N^{g_{\S}}) [\I]^{2} + \mathcal{O}([\I]^{3})$, for $g_{\I} = 1$.
\end{itemize}
Thus, in each of the three cases, the slopes of the branching, endemic equilibrium are given as
\begin{itemize}
    \item $N$, for $g_{\I} \geq 3$, 
    \item $\frac{N}{1 + 2 r N^{g_{\S}+1}} = \frac{N}{1 + r g_{\I} N^{g - 1}}$, for $g_{\I} = 2$ and
    \item $\frac{N}{1 - r (g_{\S} - 1) N^{g_{\S}}} = \frac{N}{1 - r (g - 2) N^{g - 1}}$, for $g_{\I} = 1$,
\end{itemize}
so that the bifurcations are always supercritical for $g_{\I} \geq 2$ (Fig.~\ref{fig:sis-epidemic-model}(a)), while for $g_{\I} = 1$, the bifurcation is supercritical only if $r (g_{\S} - 1) N^{g_{\S}} < 1$ and subcritical otherwise (Fig.~\ref{fig:sis-epidemic-model}(b)). Note that we say that a transcritical bifurcation is supercritical if there exists a stable branch in the (physically relevant region of) phase space beyond the bifurcation point and subcritical otherwise, corresponding to a second- or first-order transition, respectively (see Appendix~\ref{sec:appendix-transcritical-bifurcation}).

Globally, for $g_{\I} \geq 2$, we find that along the nontrivial, endemic equilibrium, eventually, a double saddle-node bifurcation is forming as the recovery rate $r$ and more specifically $r g_{\I} N^{g-1}$ is increased (Fig.~\ref{fig:sis-epidemic-model}(a)). Moreover, we can estimate that at the onset of the formation of a double saddle-node bifurcation via a codimension-two cusp bifurcation
\begin{equation*}
    r g_{\I} N^{g - 1} \approx \frac{\mathrm{e}^{(g_{\I}-1) + \sqrt{g_{\I}-1}}}{\sqrt{g_{\I}-1} ((g_{\I}-1) + \sqrt{g_{\I}-1})^{g_{\I}-2}} \, g^{g_{\I} - 2}
\end{equation*}
assuming that $g$ is large with $g \approx g_{\S} \gg g_{\I}$ (Fig.~\ref{fig:sis-epidemic-model}(c); see Appendix~\ref{sec:appendix-SIS-double-saddle-node}). For $g_{\I}$ fixed, this means that as the group size is increased, the \enquote{critical} recovery rate $r$ decreases exponentially in $g$. In turn, this means that for groups large enough, the double saddle-node bifurcation is always observed. In contrast, for $g_{\I} = 1$, we find that along the nontrivial, endemic equilibrium a (single) saddle-node bifurcation forms as soon as the transcritical bifurcation turns subcritical (Fig.~\ref{fig:sis-epidemic-model}(b)).

Recall that in the context of society, we may interpret a large recovery-promoting social component consisting of many susceptible and only a few infected individuals in terms of the influence of a functioning health system. For $g_{\I}\geq 2$, we see that this structure effectively builds up a shielding region with a very slowly growing endemic state (Fig.~\ref{fig:sis-epidemic-model}(a)). In the case of $g_{\I} = 1$, the larger the recovery rate $r$ or the group size $g$, the wider the range in which the disease-free equilibrium is stable before the first bifurcation point. In the case of $g_{\I} \geq 2$, the situation is somewhat similar: Even though the bifurcation point occurs at the same point as without the higher-order component, due to the double saddle-node bifurcation, there also exists a bistability region, although it is considerably smaller than in the previous case. However, with the disease-free equilibrium being unstable now, small outbreaks will not die out anymore but lead to an endemic equilibrium, although a relatively small one. Hence, in all cases, polyadic multiscale interactions beyond small groups can have crucial effects on epidemic dynamics. In our context, the epidemic spreading from individual to individual is counter-balanced by large group society-level effects, effectively delaying the transition to serious epidemic outbreaks to higher infection rates.

\subsection{Model II: Multiscale polyadic adaptive voter dynamics}
The voter model is a special kind of spin system and one of the classical models of opinion and consensus formation~\cite{clifford1973model,holley1975ergodic,liggett1999stochastic}. This model has been extended in various ways to account for features of realistic decision-making~\cite{redner2019reality}, one of which is the homophily, which led to what has become the adaptive voter model~\cite{holme2006nonequilibrium,kimura2008coevolutionary,zschaler2012adaptive}.
More precisely, in this model, one assumes a population of $N$ individuals, who subscribe to one of two mutually exclusive, opposing opinions, $\A$ or $\B$, and are connected via a network that encodes an underlying social network. Pairs of individuals that are connected by an edge and have opposing opinions start interacting at times given by a Poisson process. Upon interaction, with probability $1 - p$, either one of the two individuals adopts the opinion of the other (\enquote{propagation}), or with probability $p$, they break off their connection and one of the two individuals is connecting to another individual with either the same opinion (\enquote{rewire-to-same}) or any opinion (\enquote{rewire-to-random}) instead (\enquote{rewiring}/\enquote{adaptation})~\cite{zschaler2012adaptive}.
Importantly, while below a threshold probability, the topology of the network remains connected with a nonvanishing number of connected pairs of individuals with opposite opinions, beyond, the network undergoes fragmentation and the number of connected pairs of individuals with opposite opinions becomes vanishing~\cite{durrett2012graph}.

Assuming an underlying network of size $N$ with $M$ edges and thus average degree $\langle{k}\rangle = \frac{2M}{N}$, this model's mean-field evolution equations up to order 2 are given as
\begin{widetext}
\begin{equation*}
    \left\lbrace
    \begin{aligned}
        \dot{[\A]} &= (1-p) (\eta_{\A} - \eta_{B}) [\A\B] \\
        \dot{[\B]} &= (1-p) (\eta_{\B} - \eta_{A}) [\A\B] \\
        \dot{[\A\A]} &= p \frac{\pi_{\A\to\A}}{2} [\A\B] + (1-p) \eta_{\A} ([\A\B] + 2 [\A\B\A] - [\A\A\B]) \\
        \dot{[\B\B]} &= p \frac{\pi_{\B\to\B}}{2} [\A\B] + (1-p) \eta_{\B} ([\A\B] + 2 [\B\A\B] - [\B\B\A]) \\
    \end{aligned}
    \right. \quad \text{with $[\A\B] = M - ([\A\A]+[\B\B])$}
\end{equation*}
\end{widetext}
where $\pi_{\X\to\X'}$ and $\eta_{\X}$ are the probabilities of an $\X$-individual to connect to another $\X'$-individual upon rewiring and of an $\X$-individual propagating their opinion in an active edge, i.e., one connecting pairs of individuals with opposite opinions~\cite{golovin2024polyadic}. Since obviously $[\A] + [\B]$ is conserved under the dynamics, instead of considering both the evolution of $[\A]$ and $[\B]$, it turns out to be convenient to introduce as an observable their relative difference, $\mu := \frac{[\B] - [\A]}{N}$, which corresponds to the so-called magnetization of the system, a term that originated in the context of spin-systems~\cite{sood2005voter}. With that, the evolution equations are given as
\begin{widetext}
\begin{equation*}
    \left\lbrace
    \begin{aligned}
        \dot{\mu} &= (1-p) \frac{2 (\eta_{\B} - \eta_{A})}{N} [\A\B] \\
        \dot{[\A\A]} &= p \frac{\pi_{\A\to\A}}{2} [\A\B] + (1-p) \eta_{\A} ([\A\B] + 2 [\A\B\A] - [\A\A\B]) \\
        \dot{[\B\B]} &= p \frac{\pi_{\B\to\B}}{2} [\A\B] + (1-p) \eta_{\B} ([\A\B] + 2 [\B\A\B] - [\B\B\A]) \\
    \end{aligned}
    \right. \quad \text{with $[\A\B] = M - ([\A\A]+[\B\B])$}
\end{equation*}
\end{widetext}
In the case of rewire-to-random adaptation and unbiased propagation, $\pi_{\X\to\X'} = \frac{[\X']}{N}$ and $\eta_{\X} = \frac{1}{2}$, so that the magnetization is conserved and the equations reduce to the known ones~\cite{zschaler2012adaptive}. Moreover, it is known that this system may exhibit a (supercritical) transcritical bifurcation in the expected number of active edges, $[\A\B]$, where beyond the bifurcation point only the trivial equilibrium $[\A\B] = 0$ is stable and within the physical range, $0 \leq [\A\B] \leq M$, whereas below, this equilibrium is unstable and there exists a nontrivial one that is stable~\cite{zschaler2012adaptive}.

\begin{figure}[tbhp]
    \centering
    
    \includegraphics[page=3]{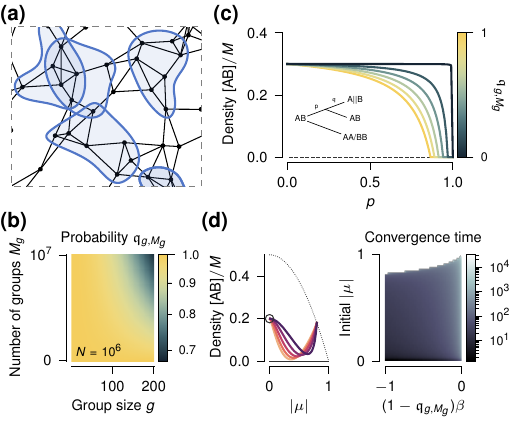}
    
    \vspace{-0.5cm}
    
    \caption{\textbf{The adaptive voter model in the presence of large groups.} \textbf{(a)}: Schematic of the local topology: The vertices of an underlying network are divided into $M_{g}$ groups of size $g$. \textbf{(b)}: The probability that a randomly chosen (2-)edge is contained in none of the $M_{g}$ $g$-hyperedges, $\mathfrak{q}_{g,M_{g}}$, is decreasing in both $M_{g}$ and $g$. \textbf{(c)}: When the containment in a hyperedge prevents the rewiring of an active edge, as $\mathfrak{q}_{g,M_{g}}$ decreases, i.e., the number $M_{g}$ or size $g$ of groups increases, the transcritical bifurcation point $p_{*}$ moves to the right ($\langle{k}\rangle = 5$, $\mu = -0.5$), stabilizing the topology. \textbf{(d)}: For that same uniform initial conditions (dotted line), the dynamics follow different trajectories depending on the value $(1-\mathfrak{q}_{g,M_{g}}) \beta$ (colored lines) to the nontrivial equilibrium (left; $\langle{k}\rangle = 5$, $p = 0.8$, $\mu_{0} = 0.8$, $(1-\mathfrak{q}_{g,M_{g}}) \beta \in \{-1.0, -0.8, \ldots -0.2\}$). When the containment in a hyperedge introduces a bias in the propagation along an active hyperedge, as $\mathfrak{q}_{g,M_{g}}$ decreases, the convergence to the nontrivial equilibrium is accelerated (right; $\langle{k}\rangle = 5$, $p = 0.8$, $\epsilon = 10^{-2}$). For initial $|\mu|$ large and $(1-\mathfrak{q}_{g,M_{g}}) \beta$ small, the dynamics do not converge to the nontrivial equilibrium (see Appendix~\ref{sec:appendix-AVM-bistability}). The numerical integration was performed using SciPy's LSODA integrator~\cite{virtanen2020scipy}.}
    \label{fig:adaptive-voter-model}
\end{figure}

In order to introduce polyadic interactions in this model in the following, we will assume that, in addition to their pairwise connections to others, the individuals are divided into groups via a set of $M_{g}$ hyperedges of size $g$ ($M_{g} \geq 1$, $g > 2$). In particular, we assume that the assignment of individuals to these groups occurs uniformly at random (and thus, is not necessarily exclusive) and remains static (Fig.~\ref{fig:adaptive-voter-model}(a)).
Under these assumptions, the probability that a given set of $\nu$ vertices is contained in at least one of the $M_{g}$ $g$-hyperedges is
\begin{equation*}
    1 - \left(1 - \frac{\binom{N-\nu}{g-\nu}}{\binom{N}{g}}\right)^{M_{g}}
\end{equation*}
which, in the limit $N \to \infty$, is nonvanishing only if $M_{g} = \Omega(N^{\nu})$, i.e., $M_{g}$ grows at least as fast as $N^{\nu}$. In particular, let $\mathfrak{q}_{g,M_{g}}$ be the probability that a randomly chosen (2-)edge is contained in none of the hyperedges,
\begin{equation*}
    \mathfrak{q}_{g,M_{g}} = \left(1 - \frac{\binom{N-2}{g-2}}{\binom{N}{g}}\right)^{M_{g}} \approx \mathrm{e}^{- \frac{M_{g}}{N^{2}} g (g-1)}.
    \label{eq:probability-q}
\end{equation*}
We can use this probability to quantify the density of hyperedges. While it is decreasing in both the number $M_{g}$ and the size $g$ of the hyperedges, we emphasize that, in contrast to the former, which enters linearly, the latter enters quadratically (Fig.~\ref{fig:adaptive-voter-model}(b)).

In the adaptive voter model, there are two competing processes, propagation and adaptation, at play; the polyadic interactions may affect either. Thus, in the following, we will consider two scenarios in which we let the presence of the additional groups given by hyperedges affect, first, the adaptation and, second, the propagation, and then investigate how this changes the overall dynamics.

In the first scenario, suppose that the hyperedges promote cohesion so that containment in a hyperedge has a stabilizing effect on the topology, meaning that if an active edge is contained in at least one of the $g$-hyperedges, the edge cannot be rewired. Consequently, by definition, a rewiring event therefore succeeds with probability $\mathfrak{q}_{g,M_{g}}$. Overall, upon interaction, rewiring will occur with probability $\mathfrak{q}_{g,M_{g}} \, p$ and propagation with probability $1 - p$ (Fig.~\ref{fig:adaptive-voter-model}(c), inset).

The mean-field evolution equations after applying the standard moment closures ($[\A\A\B] \approx \frac{2}{[\A]} [\A\A] [\A\B]$, $2 [\A\B\A] \approx \frac{1}{[\B]} [\A\B]^{2}$, $2 [\B\A\B] \approx \frac{1}{[\A]} [\A\B]^{2}$, $[\A\B\B] \approx \frac{2}{[\B]} [\A\B] [\B\B]$), also known as the pair approximation~\cite{zschaler2012adaptive,demirel2014momentclosure}, and using that $[\A] = \frac{1-\mu}{2} N$ and $[\B] = \frac{1+\mu}{2} N$ are then given as
\begin{widetext}
\begin{equation*}
    \left\lbrace
    \begin{aligned}
        \dot{[\A\A]} &= \frac{1}{2} \left(\mathfrak{q}_{g,M_{g}} \, p \, \frac{1-\mu}{2} + (1-p) \left(1 + \frac{[\A\B]}{\frac{1+\mu}{2} N} - 2\frac{[\A\A]}{\frac{1-\mu}{2} N}\right)\right) [\A\B] \\
        \dot{[\B\B]} &= \frac{1}{2} \left(\mathfrak{q}_{g,M_{g}} \, p \, \frac{1+\mu}{2} + (1-p) \left(1 + \frac{[\A\B]}{\frac{1-\mu}{2} N} - 2\frac{[\B\B]}{\frac{1+\mu}{2} N}\right)\right) [\A\B]
    \end{aligned}
    \right. \quad \text{with $[\A\B] = M - ([\A\A]+[\B\B])$.}
    \label{eq:adaptive-voter-model-1_closed}
\end{equation*}
\end{widetext}
It can be shown~\cite[Prop.~4]{kuehn2024preserving} that this system exhibits a transcritical bifurcation with nontrivial equilibrium for the active edges given as
\begin{equation*}
    [\A\B] = \frac{1 - \mu^{2}}{2} \left(1 - \frac{1 - p (1 - \frac{1 + \mu^{2}}{2} \mathfrak{q}_{g,M_{g}})}{1 - p} \frac{1}{\langle{k}\rangle} \right) M
\end{equation*}
that exists below the bifurcation point $p_{*}$ satisfying
\begin{equation*}
    \frac{1}{p_{*}} = 1 + \frac{1+\mu^{2}}{2(\langle{k}\rangle-1)} \mathfrak{q}_{g,M_{g}} .
\end{equation*}
Since $\mathfrak{q}_{g,M_{g}}$ is decreasing in both $M_{g}$ and $g$, the larger either parameter, the higher is the bifurcation point (Fig.~\ref{fig:adaptive-voter-model}(c)).
Hence, given sufficiently many and/or large hyperedges, the bifurcation point can be shifted arbitrarily close to $1$, and thus, the system is stabilized to not undergo fragmentation.

In the context of a society where individuals are frequently not on their own but are embedded into families or friendship groups, this suggests that such groups are essential for a functioning, i.e., not fragmented, society.

We also note that the order of the \enquote{decisions} being made is crucial in this scenario: If instead, before deciding on either rewiring or propagation, it is first determined whether an active edge is contained in at least one of the $g$-hyperedges, in some way, the effect becomes even stronger. This is because in this case, the probability $\mathfrak{q}_{g,M_{g}}$ appears as an overall scaling parameter for the probability $p$, so that upon interaction, rewiring will occur with probability $\mathfrak{q}_{g,M_{g}} \, p$ and propagation with probability $1 - \mathfrak{q}_{g,M_{g}} \, p$, effectively allowing to make the probability for a rewiring event arbitrarily small and this way to shift the bifurcation point beyond the physical range.

In the second scenario, suppose that the containment in a hyperedge has a destabilizing effect on the opinion landscape through introducing a bias in the propagation~\cite{redner2019reality}, meaning that in an active edge contained in at least one of the $g$-hyperedges, $\A$ and $\B$ propagate their opinion with probability $\frac{1}{2} (1-\beta\mu)$ and $\frac{1}{2} (1+\beta\mu)$, respectively, where $\beta$ interpolates the strength of the bias between biased in favor of the majority ($\beta = +1$) and in favor of the minority ($\beta = -1$) opinion. Consequently, assuming that $\beta \neq 0$, we have that the propagation probabilities are given as $\eta_{\A} = \frac{1}{2} (1-(1-\mathfrak{q}_{g,M_{g}}) \beta \mu)$ and $\eta_{\B} = \frac{1}{2} (1+(1-\mathfrak{q}_{g,M_{g}}) \beta \mu)$.

As before, the mean-field evolution equations after applying the standard moment closures are then given as
\begin{widetext}
\begin{equation*}
    \left\lbrace
    \begin{aligned}
        \dot{\mu} &= (1-p) \frac{2 (1-\mathfrak{q}_{g,M_{g}}) \beta}{N} \mu \, [\A\B] \\
        \dot{[\A\A]} &= \frac{1}{2} \left(p \, \frac{1-\mu}{2} + (1-p) (1-(1-\mathfrak{q}_{g,M_{g}}) \beta \mu) \left(1 + \frac{[\A\B]}{\frac{1+\mu}{2} N} - 2\frac{[\A\A]}{\frac{1-\mu}{2} N}\right)\right) [\A\B] \\
        \dot{[\B\B]} &= \frac{1}{2} \left(p \, \frac{1+\mu}{2} + (1-p) (1+(1-\mathfrak{q}_{g,M_{g}}) \beta \mu) \left(1 + \frac{[\A\B]}{\frac{1-\mu}{2} N} - 2\frac{[\B\B]}{\frac{1+\mu}{2} N}\right)\right) [\A\B]
    \end{aligned}
    \right. \quad \text{with $[\A\B] = M - ([\A\A]+[\B\B])$}
    \label{eq:adaptive-voter-model-2_closed}
\end{equation*}
\end{widetext}
where we immediately note that the magnetization is no longer conserved.

However, provided that $\beta \leq 0$, it can be shown~\cite[Prop.~4]{kuehn2024preserving} that this system also exhibits a transcritical bifurcation. Moreover, at equilibrium, we necessarily have that $\mu = 0$ or $[\A\B] = 0$. Hence, a nontrivial equilibrium for the active edges can only exist if $\mu = 0$ in which case it is given as 
\begin{equation*}
    [\A\B] = \frac{1}{2} \left(1 - \frac{1 - \frac{1}{2} p}{1 - p} \frac{1}{\langle{k}\rangle} \right) M
\end{equation*}
below the transcritical bifurcation point $p_{*}$ satisfying
\begin{equation*}
    \frac{1}{p_{*}} = 1 + \frac{1}{2(\langle{k}\rangle-1)} .
\end{equation*}
Importantly, we observe that the nontrivial equilibrium can only ever exist if $\beta \leq 0$.

While the existence of the $g$-hyperedges has no impact on the equilibrium and the bifurcation, they ensure the propagation of the bias in the sense that they accelerate convergence to equilibrium. This can be seen considering the time it takes a trajectory starting from uniform initial conditions ($[\A\A] = \frac{1}{4} (1 - \mu)^{2} M$ and $[\B\B] = \frac{1}{4} (1 + \mu)^{2} M$) to reach some $\epsilon$-neighborhood around the nontrivial equilibrium point for $p < p_{*}$ (Fig.~\ref{fig:adaptive-voter-model}(d)). It should be noted, though, that the equilibrium is not globally attracting. Even though it exists stably, trajectories starting close to a global consensus state may actually converge to a trivial equilibrium (see Appendix~\ref{sec:appendix-AVM-bistability}).

Again in the context of a society, this shows that even the slightest institutional (majority) mistrust, which manifests itself or is promoted within social groups and lets people preferably side with the minority, eventually leads to a deeply divided population~\cite{finkel2020political}. In fact, such a polarization has been identified as only one of several negative societal outcomes of institutional mistrust~\cite{vanprooijen2022suspicion}. While the society is still functioning, in the sense that it is not fragmented, there is no (stable) majority for one opinion or the other anymore, complicating government.

\section{Discussion}
In this work, we have studied multiscale polyadic interactions, where small group effects interact with very large population-level coupling. We have derived mean-field models of an epidemic and an adaptive voter multiscale model. The main result of the analysis of these models is that large-group effects cannot be neglected, as they may counteract small-group interactions. In the case of the epidemic model, while the phase transition toward an endemic equilibrium still exists, it can be lowered by large group recovery-type interactions, or large groups induce a shielding effect through the emergence of a double saddle-node bifurcation with a very flat endemic branch near the disease-free state. In the case of the adaptive voter model, the known phase transition remains qualitatively the same but shifts as the group sizes are increased. In particular, we have shown how large groups may have a stabilizing effect on the topology, preventing the population from fragmentation.

Throughout this work, we have always assumed that polyadic interactions occur within groups of a fixed, absolute size that is significantly larger than $2$ but independent of the overall population size. In fact, this setting is motivated by the fact that there is an (absolute) limit to the number of relationships humans can maintain, which suggests that there is a limit on the size of social groups irrespective of the overall population size~\cite{dunbar1992neocortex,saramaki2014persistence}. Similarly, it has been reported that there is an (economically) ideal size to societal support structures like hospitals~\cite{posnett1999bigger,giancotti2017efficiency}.

Overall, within many further complex systems, we conjecture that multiscale polyadic dynamics, i.e., a scale separation between small and large groups is crucial. An immediate example is economic dynamics (e.g., games on hypergraphs), where small-scale direct economic interaction is coupled to a large-scale globalized economy.


\subsubsection*{Acknowledgements}
P.S. and C.K. acknowledge funding from the Deut\-sche For\-schungs\-ge\-mein\-schaft (DFG, German Research Foundation) via SPP 2265: \enquote{Random Geometric Systems} (Project 443731539). J.M. and C.K. acknowledge funding from the Volks\-wagen\-Stif\-tung (Volks\-wagen Foundation) via a Lich\-ten\-berg Professorship awarded to C.K.


\appendix
\section{Supplementary Information}
\label{sec:appendix}

\renewcommand{\thefigure}{A\arabic{figure}}
\setcounter{figure}{0}

\subsection{Simplicial network moment closure}
\label{sec:appendix-simplicial-closure}

Suppose that in a simplicial $(N-1)$-network, individuals are in states $\S$ or $\I$ with probability $1 - p$ and $p$ so that $[\S] = (1-p) N$ and $[\I] = p N$. In particular, the number of individuals in state $\I$ is binomially distributed, $N_{\I} \sim \operatorname{Bin}(N,p)$.

Consequently, with probability $\mathbb{P}[\operatorname{Bin}(N,p) = N_{\I}]$, there are $g_{\S}! g_{\I}! \binom{N_{\S}}{g_{\S}} \binom{N_{\I}}{g_{\I}}$ simplices with $g_{\S}$ and $g_{\I}$ (ordered) individuals in states $\S$ and $\I$, respectively, for $N_{I} = g_{\I}, \ldots N - g_{\S}$ ($N_{\S} = N - N_{\I}$). Hence, in expectation, their number is
\begin{equation*}
    \begin{aligned}
        &\phantom{=} g_{\S}! g_{\I}! \sum_{N_{I} = g_{I}}^{N - g_{\S}} \binom{N_{\S}}{g_{\S}} \binom{N_{\I}}{g_{\I}}  \binom{N}{N_{\I}} (1-p)^{N_{\S}} p^{N_{\I}} \\
        &= \frac{N!}{(N-g)!} (1-p)^{g_{\S}} p^{g_{\I}} \\
        &= \frac{1}{N^{g}} \frac{N!}{(N-g)!} [\S]^{g_{\S}} [\I]^{g_{\I}}
    \end{aligned}
\end{equation*}
where in the last step, we wrote $1-p$ and $p$ in terms of $[\S]$ and $[\I]$.

For $N \gg g$, one has that $\frac{1}{N^{g}} \frac{N!}{(N-g)!} \approx 1$ so that we conclude that
\begin{equation*}
    [\triangle\lbrace \S^{g_{\S}} \I^{g_{\I}} \rbrace] \approx [\S]^{g_{\S}} [\I]^{g_{\I}} .
\end{equation*}

Note that this moment closure does hold regardless of whether $[\I] < g_{\I}$ or $[\I] \geq g_{\I}$ and, analogously, $[\S] < g_{\S}$ or $[\S] \geq g_{\S}$. Indeed, one cannot naively argue that whenever $[\X] < g_{\X}$, $[\triangle\lbrace \S^{g_{\S}} \I^{g_{\I}} \rbrace] = 0$. This is because even though statistically $[\X] < g_{\X}$, as per the above argument, there are always realizations for which there are more than $g_{X}$ individuals in state $\X$ so that necessarily $[\triangle\lbrace \S^{g_{\S}} \I^{g_{\I}} \rbrace] > 0$.

\subsection{Normal forms of a transcritical bifurcation in scalar dynamical systems}
\label{sec:appendix-transcritical-bifurcation}

The mean-field evolution equation for the SIS-epidemic model, in a neighborhood of $0$, is of the form
\begin{equation*}
    \dot{[\I]} = (p - p_{*}) [\I] - a [\I]^{2} + \mathcal{O}([\I]^{3}).
\end{equation*}
Then, upon rescaling time (note the different symbol to denote the (time-)derivative), the system is equivalent to
\begin{equation*}
    [\I]' = \frac{p - p_{*}}{a} [\I] - [\I]^{2} + \mathcal{O}([\I]^{3}).
\end{equation*}

This is readily recognized as the normal form of a transcritical bifurcation~\cite{strogatz2024nonlinear}. Hence, $[\I] = 0$ is an equilibrium and undergoes a transcritical bifurcation at $p = p_{*}$. In particular, this equilibrium is stable below and unstable above the bifurcation point. More specifically, if the branching, nontrivial equilibrium in the physically relevant region of phase space, $[0, N]$, is stable, we call the bifurcation supercritical, and if it is unstable, subcritical~\cite{simonnet2009bifurcation,kuehn2021universal,kuehn2024preserving}. In particular, since the slope of the branching equilibrium is given as $\frac{1}{a}$, the bifurcation is supercritical if $a > 0$ and subcritical if $a < 0$. 

\subsection{Formation of the double saddle-node bifurcation in the SIS-epidemic model}
\label{sec:appendix-SIS-double-saddle-node}

For the the SIS-epidemic model, the equilibria $(\lambda N, \frac{[\I]}{N})$ along the nontrivial bifurcation branch can be parameterized as
\begin{equation*}
     \left( \frac{1}{1 - \xi} + r g_{\I} N^{g - 1} (1 - \xi)^{g_{\S} - 1} \xi^{g_{\I} - 1}, \xi \right) \quad \text{for $\xi \in [0,1]$.}
\end{equation*}

Hence, it is straightforward to verify that the formation of a local maximum (and minimum) in the graph of the function $\frac{1}{1 - \xi} + r g_{\I} N^{g - 1} (1 - \xi)^{g_{\S} - 1} \xi^{g_{\I} - 1}$ in the unit interval as the value of $r g_{\I} N^{g - 1}$ is increased for $g_{\S}$ and $g_{\I}$ fixed is a necessary condition for the formation of a saddle-node and then a double saddle-node bifurcation along this branch.
With $\gamma := r g_{\I} N^{g - 1}$ and assuming that $g_{\S}, g_{I} \geq 2$, this is the case if $\tfrac{\mathrm{d}}{\mathrm{d}\xi} (\frac{1}{1 - \xi} + \gamma (1 - \xi)^{g_{\S} - 1} \xi^{g_{\I} - 1}) \propto \xi^{2} - \gamma (g - 2) (1 - \xi)^{g_{\S}} \xi^{g_{\I}} (\xi - \frac{g_{\I} - 1}{g - 2}) =: \xi^{2} - \gamma \phi_{(g_{\S},g_{\I})}(\xi) = 0$ for some $0 < \xi < 1$.

\begin{figure}[htbp]
    \centering
    
    \includegraphics[page=2]{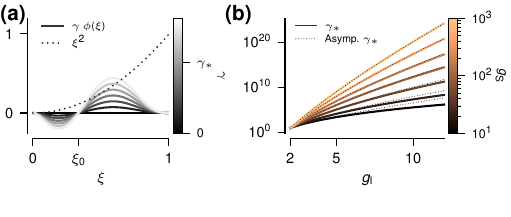}

    \vspace{-0.5cm}
    
    \caption{\textbf{Formation of the double saddle-node bifurcation.} \textbf{(a)}: The existence of a solution to $\xi^{2} - \gamma \phi_{(g_{\S},g_{\I})}(\xi) = 0$ for $0 < \xi < 1$ (or rather $\xi_{0} < \xi < 1$) characterizes the presence of double saddle-node bifurcation ($g_{\S} = 3$, $g_{\I} = 2$). As $\gamma$ is increased, solutions eventually exist. Hence, the minimal value for $\gamma$, $\gamma_{*}$, such that a solution exists characterizes its formation. \textbf{(b)}: In comparison with the true critical values $\gamma_{*}$, the asymptotic value provides an excellent estimate for $g_{\S}$ sufficiently large.}
    \label{fig:sis-epidemic-model-supp}
\end{figure}

$\phi_{(g_{\S},g_{\I})}$ is a polynomial in $\xi$ with zeros $0$, $\frac{g_{\I} - 1}{g - 2} =: \xi_{0}$, and $1$, where $0 < \xi_{0} < 1$, and since $\phi_{(g_{\S},g_{\I})}'(\xi_{0}) = (g-2) (1 - \xi_{0})^{g_{\S}} \xi_{0}^{g_{\I}} > 0$, $\phi_{(g_{\S},g_{\I})}(\xi) < 0$ for $0 < \xi < \xi_{0}$ and $\phi_{(g_{\S},g_{\I})}(\xi) > 0$ for $\xi_{0} < \xi < 1$. Hence, for $\xi^{2} - \gamma \phi_{(g_{\S},g_{\I})}(\xi) = 0$ to hold, necessarily $\xi_{0} < \xi < 1$. For $\gamma = 0$, there exists no such solution, while there exist up to two solutions if $\gamma$ is sufficiently large (Fig.~\ref{fig:sis-epidemic-model-supp}(a)). We let $\gamma_{*}$ be such that $\xi^{2} - \gamma_{*} \phi_{(g_{\S},g_{\I})}(\xi) = 0$ has exactly one solution for $\xi_{0} < \xi < 1$. In particular, we note that the value of $\gamma_{*}$ characterizes the onset of the formation of the double saddle-node bifurcation.

For $\xi^{2} - \gamma_{*} \phi_{(g_{\S},g_{\I})}(\xi) = 0$ to have only one solution for $\xi_{0} < \xi < 1$, we necessarily also require that $2 \xi - \gamma_{*} \phi'_{(g_{\S},g_{\I})}(\xi) = 0$. Together, this yields that $\frac{2}{\xi} = \tfrac{\mathrm{d}}{\mathrm{d}\xi} \ln{\phi_{(g_{\S},g_{\I})}(\xi)}$ whose only solution for $\xi_{0} < \xi < 1$ is
\begin{equation*}
    \xi_{*} = \frac{g_{\I}-1}{g-1} \left(1 + \sqrt{1 - \frac{g-1}{g_{\I}-1} \frac{g_{\I}-2}{g-2}} \right) .
\end{equation*}
Hence, $\gamma_{*} = \frac{\xi_{*}^{2}}{\phi_{(g_{\S},g_{\I})}(\xi_{*})}$ so that if
\begin{equation*}
    r g_{\I} N^{g - 1} \equiv \gamma > \gamma_{*} = \frac{\xi_{*}^{2}}{\phi_{(g_{\S},g_{\I})}(\xi_{*})},
    \label{eq:double-saddle-node-threshold-value}
\end{equation*}
the nontrivial bifurcation branch contains a double saddle-node bifurcation.

As for asymptotic behavior for $g \to \infty$ with $g_{\I}$ fixed, we have that $\xi_{*} \sim \frac{(g_{\I}-1) + \sqrt{g_{\I}-1})}{g}$ and thus
\begin{equation*}
    \gamma_{*} \sim \frac{\mathrm{e}^{(g_{\I}-1) + \sqrt{g_{\I}-1}}}{\sqrt{g_{\I}-1} ((g_{\I}-1) + \sqrt{g_{\I}-1})^{g_{\I}-2}} \, g^{g_{\I} - 2}.
    \label{eq:double-saddle-node-threshold-value_asymptotics}
\end{equation*}

%
%
%

\subsection{Emergence of a bistability region in the adaptive voter model with bias}
\label{sec:appendix-AVM-bistability}

A rarely discussed feature of the adaptive model (with or without bias) is the fact that the nontrivial equilibrium may not be globally attracting. In these cases, subsets within the manifold of trivial equilibria are attracting, which leads to the emergence of a region of bistability. Depending on the initial conditions, it can happen that even though a locally stable nontrivial equilibrium exists, the system approaches a trivial equilibrium.

To illustrate this, let $\eta_{\A} = \frac{1}{2} (1-\hat{\beta} \mu)$ and $\eta_{\B} = \frac{1}{2} (1+\hat{\beta} \mu)$, which for $\hat{\beta} = 0$ and $\hat{\beta} \neq 0$ yields the adaptive voter model without or with bias, respectively, and the mean-field evolution equations after applying the standard moment are then given as
\begin{widetext}
\begin{equation*}
    \left\lbrace
    \begin{aligned}
        \dot{\mu} &= (1-p) \frac{2 \hat{\beta}}{N} \mu \, [\A\B] \\
        \dot{[\A\A]} &= \frac{1}{2} \left(p \, \frac{1-\mu}{2} + (1-p) (1-\hat{\beta} \mu) \left(1 + \frac{[\A\B]}{\frac{1+\mu}{2} N} - 2\frac{[\A\A]}{\frac{1-\mu}{2} N}\right)\right) [\A\B] \\
        \dot{[\B\B]} &= \frac{1}{2} \left(p \, \frac{1+\mu}{2} + (1-p) (1+\hat{\beta} \mu) \left(1 + \frac{[\A\B]}{\frac{1-\mu}{2} N} - 2\frac{[\B\B]}{\frac{1+\mu}{2} N}\right)\right) [\A\B]
    \end{aligned}
    \right. \quad \text{with $[\A\B] = M - ([\A\A]+[\B\B])$.}
    \label{eq:biased-adaptive-voter-model_closed}
\end{equation*}
\end{widetext}

The physically relevant region of phase space with coordinates $(\mu, [\A\A], [\B\B])$ is given by the simplicial cylinder $[-1,+1] \times \Delta^{(2)}_{M}$ where $\Delta^{(2)}_{M} := \lbrace x \in \mathbb{R}^{2} : x \geq 0 \wedge \Vert{x}\Vert_{1} \leq M \rbrace$ is a $2$-simplex with \enquote{radius} $M$. In particular, we note that the trivial equilibria form the manifold $[-1,+1] \times \partial \Delta^{(2)}_{M}$ where $\partial \Delta^{(2)}_{M} := \lbrace x \in \mathbb{R}^{2} : x \geq 0 \wedge \Vert{x}\Vert_{1} = M \rbrace$ and we parametrize the points of the trivial equilibria as $(\mu, (1 - \theta)M, \theta M)$ for magnetization $-1 \leq \mu \leq +1$ and an interpolation parameter $0 \leq \theta \leq 1$.

\begin{figure}[htbp]
    \centering
    
    \includegraphics[page=4]{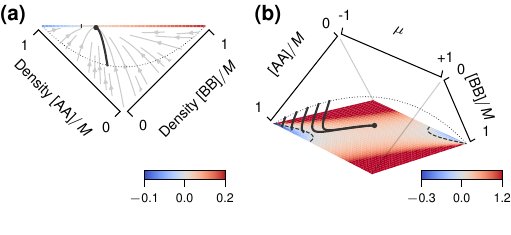}

    \vspace{-0.5cm}
    
    \caption{\textbf{Emergence of a bistability region in the adaptive voter model.} Sufficiently close to the bifurcation point, along the manifold of trivial equilibria, (linear) stability changes as indicated by the value of the only nonvanishing eigenvalue of the linearization (colorbar). Consequently, it may happen that both the nontrivial and some subset of the trivial equilibria are stable. The set of uniform initial conditions forms a parabola in the phase space (dotted line). \textbf{(a)}: Phase space of the adaptive voter model without bias and fixed magnetization. For uniform initial conditions, below the bifurcation point, the dynamics converge to the nontrivial equilibrium ($\langle{k}\rangle = 5$, $p = 0.88$, $\mu_{0} = -0.2$). The numerical integration was performed using SciPy's LSODA integrator~\cite{virtanen2020scipy}. \textbf{(b)}: Full phase space of the adaptive voter model with bias. For different uniform initial conditions (dotted line) and below the bifurcation point, the dynamics converge either to a trivial or the nontrivial equilibrium ($\langle{k}\rangle = 5$, $p = 0.88$, $\hat{\beta} = -0.2$, $\mu_{0} \in \{-0.9, -0.8, \ldots -0.5\}$). The numerical integration was performed using SciPy's LSODA integrator~\cite{virtanen2020scipy}.}
    \label{fig:adaptive-voter-model-supp}
\end{figure}

In terms of stability, we have that a trivial equilibrium $(\mu, (1 - \theta)M, \theta M)$ is (linearly) stable if
\begin{equation*}
    -\frac{(1-\hat{\beta}) \mu (\mu - (2 \theta - 1))}{1 - \mu^{2}} > 1 - \frac{1 - \frac{1}{2} p}{1-p} \frac{1}{\langle{k}\rangle} .
\end{equation*}
Due to the degeneracy of the trivial equilibria, all but one eigenvalue of the linearization on the manifold vanish~\cite{kuehn2024preserving}, so this condition ensures that the only nonvanishing eigenvalue is negative.

If there is no bias, i.e., $\hat{\beta} = 0$, this reduces to the condition
\begin{equation*}
    -\frac{\mu (\mu - (2 \theta - 1))}{1 - \mu^{2}} > 1 - \frac{1 - \frac{1}{2} p}{1-p} \frac{1}{\langle{k}\rangle}
\end{equation*}
for the interpolation parameter $\theta$ since the magnetization is conserved in this case.

In fact, one finds that, especially close to the bifurcation point, there are values $\theta$ such that the condition is satisfied. Consequently, just before the bifurcation point, it can happen that subsets within the manifold of trivial equilibria are stable, while also the nontrivial equilibrium exists and is stable (Fig.~\ref{fig:adaptive-voter-model-supp}(a)). It should be noted, though, that these subsets do depend on the magnetization and, since it is conserved, the initial conditions. Hence, for fixed magnetization, it could be that there are no initial conditions so that the dynamics actually reach the basin of attraction of these stable subsets of trivial equilibria.

This is in contrast to the case when there is a bias, i.e., $\hat{\beta} \neq 0$, and the magnetization is not conserved anymore. Again, one finds that especially close to the bifurcation point, there are values $(\mu, \theta)$ such that the stability condition is satisfied, so that subsets within the manifold of trivial equilibria are stable, while also the nontrivial equilibrium exists and is stable (Fig.~\ref{fig:adaptive-voter-model-supp}(b)). Moreover, for $\hat{\beta} < 0$, uniform initial conditions sufficiently close to a global consensus state are actually in the basin of attraction of these stable subsets of trivial equilibria.

\vfill


\bibliography{references}

\end{document}